\documentclass[useAMS,usenatbib]{mn2e}

\usepackage{graphicx}
\usepackage{url}
\usepackage{amsmath}
\usepackage{amssymb}

\title[$f(T)$ gravity's effects and upper-bounds]{$f(T)$ gravity: effects on astronomical observation and Solar System experiments and upper-bounds}
\author[Y. Xie \& X.-M. Deng]{Yi Xie$^{1,3}$\thanks{E-mail:yixie@nju.edu.cn} and Xue-Mei Deng$^{2,3}$\\
$^{1}$Department of Astronomy, Nanjing University, Nanjing 210093, China\\
$^{2}$Purple Mountain Observatory, Chinese Academy of Sciences, Nanjing 210008, China\\
$^{3}$Key Laboratory of Modern Astronomy and Astrophysics, Nanjing University, Ministry of Education, Nanjing 210093, China}
\begin{document}

\date{Accepted . Received ; in original form }

\pagerange{\pageref{firstpage}--\pageref{lastpage}} \pubyear{2002}

\maketitle

\label{firstpage}

\begin{abstract}
As an extension of a previous work in which perihelion advances are considered only and as an attempt to find more stringent constraints on its parameters, we investigate effects on astronomical observation and experiments conducted in the Solar System due to the $f(T)$ gravity which contains a quadratic correction of $\alpha T^2$ ($\alpha$ is a model parameter) and the cosmological constant $\Lambda$. Using a spherical solution describing the Sun's gravitational field, the resulting secular evolution of planetary orbital motions, light deflection, gravitational time delay and frequency shift are calculated up to the leading contribution. Among them, we find qualitatively that the light deflection holds a unique bound on $\alpha$, without dependence on $\Lambda$, and the time delay experiments during inferior conjunction impose a clean constraint on $\Lambda$, regardless of $\alpha$. Based on observation and experiments, especially the supplementary advances in the perihelia provided by the INPOP10a ephemeris, we obtain the upper-bounds quantitatively: $|\alpha| \le  1.2 \times 10^{2}$ m${}^2$ and $ |\Lambda| \le 1.8 \times 10^{-43}$ m${}^{-2}$, at least 10 times tighter than the previous result.
\end{abstract}


\section{Introduction}           %% first-level sections will be auto-capitalized

Observations \citep[e.g.][]{Riess1998AJ116.1009, Perlmutter1999ApJ517.565} show we live in a universe undergoing an accelerating expansion. One possible way to explain it is to introduce the presence of repulsive dark energy as an odd ingredient \citep[for a recent review see][and references therein]{Amendola2010}; the other way is to modify the theory of gravity and these modified theories can generate interesting cosmological consequences \citep[for a recent review see][and references therein]{Clifton2012PhR513.1}.

A new and developing kind of modified gravity is the $f(T)$ gravity \citep{Ferraro2007PRD75.084031,Ferraro2008PRD78.124019,Bengochea2009PRD79.124019,Linder2010PRD81.127301}, which is an extension of the teleparallel equivalent of General Relativity(GR) \citep{Hayashi1979PRD19.3524}. Albert Einstein originally proposed the idea of the teleparallel, trying to unify gravitation and electromagnetism \citep[for English translation of his original papers around 1928 see][]{Unzicker2005}. Similar to the $f(R)$ theory of gravity generalizing Einstein-Hilbert action \citep[for a recent review see][and references therein]{Felice2010LRR13.3}, the $f(T)$ gravity extends earlier formalism of the teleparallel by making the Lagrangian density a function of $T$ which is a torsion scalar.

Cosmology in the framework of the $f(T)$ gravity has been intensively studied \citep[for a list of extensive literature see][]{Iorio2012MNRAS427.1555}. Cosmological observations data are taken to test the proposal that uses it to explain cosmic accelerating expansion without dark energy \citep{Wu2010PLB693.415,Bengochea2011PLB695.405}. In order to obtain tighter bounds for its parameters, \citet{Iorio2012MNRAS427.1555} obtain a spherical solution of the $f(T)$ gravity with a quadratic correction of $\alpha T^2$, where $\alpha$ is a model parameter. The solution also contains contribution due to the cosmological constant $\Lambda$ and reads as
\begin{equation}
  \label{sphsym}
  \mathrm{d}s^2 = N(r)^2 c^2\mathrm{d}t^2-K(r)^{-2}\mathrm{d}r^2-r^2\mathrm{d}\Omega^2,
\end{equation}
where $c$ is the speed of light, $r$ is the radial distance from the origin, $\mathrm{d}\Omega^2 = \mathrm{d}\theta^2+\sin^2\theta \mathrm{d}\phi^2$ and, up to $\mathcal{O}(\alpha^2/r^4)$,
\begin{eqnarray}
  \label{N^2}
  N(r)^2 & = & 1 -\frac{2GM}{c^2r} -\frac{\Lambda}{3}r^2 -6\frac{\alpha}{r^2}, \\
  \label{K^2}
  K(r)^2 & = & 1 -\frac{2GM}{c^2r} -\frac{\Lambda}{3}r^2 -14\frac{\alpha}{r^2}.
\end{eqnarray}
In above ones, the leading corrections to Schwarzschild spacetime caused by $\alpha$ and $\Lambda$ are kept only and more detailed expressions for $N(r)^2$ and $K(r)^2$ can be found in \citet{Iorio2012MNRAS427.1555}. Through describing the Sun's gravitation field by equation (\ref{sphsym}), \citet{Iorio2012MNRAS427.1555} investigate possible deviations in the secular precession of the longitude of the pericentre for planets in the Solar System and find their bounds: $|\alpha|\le 1.8 \times 10^4$ m${}^2$ and $|\Lambda|\le 6.1 \times 10^{-42}$ m${}^{-2}$.

In this work, as an extension of the previous work \citep{Iorio2012MNRAS427.1555} in which perihelion advances are considered only and as an attempt to find more stringent upper-bounds on its parameters, we investigate the $f(T)$ gravity's effects on astronomical observation and experiments conducted in the Solar System. The secular evolution of planetary orbital motions, light deflection, gravitational time delay and frequency shift are calculated up to the leading contribution.

In summary, we find qualitatively that the light deflection holds a unique bound on $\alpha$, without dependence on $\Lambda$, and the time delay experiments during inferior conjunction impose a clean constraint on $\Lambda$, regardless of $\alpha$. And, based on observation and experiments, especially the supplementary advances in the perihelia provided by the INPOP10a ephemeris \citep{Fienga2011CMDA111.363}, we obtain the upper-bounds quantitatively: $|\alpha| \le  1.2 \times 10^{2}$ m${}^2$ and $ |\Lambda| \le 1.8 \times 10^{-43}$ m${}^{-2}$, at least 10 times tighter than the previous result \citep{Iorio2012MNRAS427.1555}.

The rest of the paper is organized as follows. Section \ref{obs} is devoted to calculating the $f(T)$ gravity's effects on observation and experiments. In Section \ref{bounds}, we estimate the upper-bounds for $\alpha$ and $\Lambda$ based on observation and experiments. Finally, in Section \ref{candd}, we summarize and discuss our results.

\section{Effects on observation and experiments }

\label{obs}

Although the Solar System consists of many gravitating bodies, the dominance of the Sun makes the spherical symmetric spacetime described by equation (\ref{sphsym}) a sufficiently good approximation for modeling quite a large number of observations and experiments. In the following parts, we will stay with this spacetime and investigate the leading deviations caused by the $f(T)$ gravity from GR.

\subsection{Secular evolution of planetary orbits}

\label{perihelion}

In this part, we will focus on the long term evolution of planetary dynamics as an extension of the previous work \citep{Iorio2012MNRAS427.1555} in which perihelion shifts are considered only. For a planet around the Sun (with a bound orbit), the leading corrections on its relativistic trajectory caused by $\alpha$ and $\Lambda$ can be written as a perturbing potential \citep{Iorio2012MNRAS427.1555}
\begin{equation}
  \label{}
  R = \frac{1}{6}\Lambda c^2 r^2 +3\alpha\frac{ c^2}{r^2}
\end{equation}
where $c$ is the speed of light and $r$ is the radial distance between the planet and the Sun. Changes of its osculating orbital elements, $a$, $e$, $i$, $\Omega$, $\omega$ and $M$\footnote{With widely used notations in celestial mechanics, $a$ is the semi-major axis, $e$ is the eccentricity, $i$ is the inclination, $\Omega$ is the longitude of the ascending node, $\omega$ is the argument of periastron and $M$ is the mean anomaly.}, satisfy the Lagrange planetary equations \citep{Danby1962}
\begin{eqnarray}
\frac{\mathrm{d}a}{\mathrm{d}t}&=&\frac{2}{na}\frac{\partial R}{\partial M},\\
\frac{\mathrm{d}e}{\mathrm{d}t}&=&\frac{(1-e^{2})}{na^{2}e}\frac{\partial R}{\partial M}-\frac{\sqrt{1-e^{2}}}{na^{2}e}\frac{\partial R}{\partial\omega},\\
\frac{\mathrm{d}i}{\mathrm{d}t}&=&\frac{\cos i}{na^{2}\sqrt{1-e^{2}}\sin i}\frac{\partial R}{\partial\omega}-\frac{1}{na^{2}\sqrt{1-e^{2}}\sin i}\frac{\partial R}{\partial\Omega},\\
\frac{\mathrm{d}\Omega}{\mathrm{d}t}&=&\frac{1}{na^{2}\sqrt{1-e^{2}}\sin i}\frac{\partial R}{\partial i},\\
\frac{\mathrm{d}\omega}{\mathrm{d}t}&=&\frac{\sqrt{1-e^{2}}}{na^{2}e}\frac{\partial R}{\partial e}-\frac{\cos i}{na^{2}\sqrt{1-e^{2}}\sin i}\frac{\partial R}{\partial i},\\
\frac{\mathrm{d}M}{\mathrm{d}t}&=&n-\frac{(1-e^{2})}{na^{2}e}\frac{\partial R}{\partial e}-\frac{2}{na}\frac{\partial R}{\partial a}.
\end{eqnarray}
For secular evolution, averaging $R$ over one orbital revolution is needed and that is
\begin{equation}
   \label{}
   \bar{R}  = \frac{1}{T}\int_0^TR \mathrm{d}t = \frac{1}{6}\Lambda c^2 a^2 \bigg(1+\frac{3}{2}e^2\bigg) + 3\alpha \frac{c^2}{a^2}(1-e^2)^{-1/2}.
\end{equation}
Therefore, we can have the long term variations of the orbit as
\begin{eqnarray}
  \label{}
  \bigg<\frac{\mathrm{d} a}{\mathrm{d} t}\bigg> & = & 0,\\
  \bigg<\frac{\mathrm{d} e}{\mathrm{d} t}\bigg> & = & 0,\\
  \bigg<\frac{\mathrm{d} i}{\mathrm{d} t}\bigg> & = & 0,\\
  \bigg<\frac{\mathrm{d} \Omega}{\mathrm{d} t}\bigg> & = & 0,\\
  \label{seculardomegadt}
  \bigg<\frac{\mathrm{d} \omega}{\mathrm{d} t}\bigg> & = & \frac{\sqrt{1-e^2}}{2n}\Lambda c^2 +3\alpha \frac{c^2}{na^4(1-e^2)},\\
  \bigg<\frac{\mathrm{d} M}{\mathrm{d} t}\bigg> & = & n -\bigg(\frac{7}{6}+\frac{1}{2}e^2\bigg)\frac{\Lambda c^2}{n}\nonumber\\
  & & +9\alpha\frac{c^2}{na^4}(1-e^2)^{-1/2},
\end{eqnarray}
where (\ref{seculardomegadt}) totally agrees with the one derived by \citet{Iorio2012MNRAS427.1555}. Therefore, we can define the observed deviation from GR caused by $f(T)$ gravity as
\begin{eqnarray}
  \label{dGRomega}
  \delta^{\mathrm{GR}}_{\left<\dot{\omega}\right>} & \equiv &  |\left<\dot{\omega}\right>_{\mathrm{obs}} - \left<\dot{\omega}\right>_{\mathrm{GR}}|\nonumber\\
  & = & \bigg| \frac{\sqrt{1-e^2}}{2n}\Lambda c^2 +3\alpha \frac{c^2}{na^4(1-e^2)} \bigg|.
\end{eqnarray}

\subsection{Light deflection}

The motion of a photon in the spacetime given by equation (\ref{sphsym}) holds the relation as
\begin{equation}
  \label{}
  0 = N(r)^2 c^2\dot{t}^2 -K(r)^{-2} \dot{r}^2 -r^2\dot{\phi}^2,
\end{equation}
where the dots stand for differentiation against the affine parameter $\lambda$ and since the gravitational field is isotropic we may consider the orbit of the photon to be confined to the equatorial plane that $\theta=\pi/2$ \citep{Weinberg1972Book}. Its has two conserved quantities along the light trajectory:
\begin{equation}
  \label{EL}
  E \equiv N(r)^2 c^2\dot{t},\qquad L\equiv r^2 \dot{\phi}.
\end{equation}
They give
\begin{equation}
  \label{dphidr}
  \frac{\mathrm{d}\phi}{\mathrm{d}r} = \pm \frac{1}{r^2} \frac{N}{K} \bigg(\frac{1}{b^2}-\frac{N^2}{r^2}\bigg)^{-1/2},
\end{equation}
where $b\equiv L/E$. For the closest approach $d$, $\mathrm{d}r/\mathrm{d}\phi=0$ leads to
\begin{equation}
  \label{}
  b = \frac{d}{N(d)}.
\end{equation}
Thus, we can obtain
\begin{eqnarray}
  \label{}
  \frac{\mathrm{d}\phi}{\mathrm{d}r} & = & \frac{d}{r\sqrt{r^2-d^2}} + \frac{GM}{c^2r^2}\frac{(d^2+dr+r^2)}{(d+r)\sqrt{r^2-d^2}}\nonumber\\
  & & +\alpha \frac{(7d^2+3r^2)}{dr^2\sqrt{r^2-d^2}},
\end{eqnarray}
and
\begin{equation}
  \label{}
  \phi = 2 \int_d^{\infty} \frac{\mathrm{d}\phi}{\mathrm{d}r} \mathrm{d}r = \pi + 4\frac{GM}{c^2d} +\alpha \frac{13\pi}{2d^2}.
\end{equation}

The deflection angle can be worked out as
\begin{equation}
  \label{fTlightdef}
  \Delta\phi = 4\frac{GM}{c^2d} +\alpha \frac{13\pi}{2d^2}.
\end{equation}
A key point here is that only $\alpha$ affects the measurements of light deflection which is immune to $\Lambda$. The absence of $\Lambda$ in light bending matches previous results \citep{Lake2002PRD65.087301,Kagramanova2006PLB634.465}. The deviation in the deflection from GR is
\begin{equation}
  \label{dGRphi}
  \delta^{\mathrm{GR}}_{\Delta \phi} \equiv | \Delta \phi_{\mathrm{obs}} - \Delta \phi_{\mathrm{GR}} | = |\alpha| \frac{13\pi}{2d^2}.
\end{equation}

\subsection{Gravitational time delay}

From equations (\ref{EL}) and (\ref{dphidr}), we can obtain the relationship between $t$ and $r$ for light as
\begin{equation}
  \label{dtdr}
  \frac{\mathrm{d}t}{\mathrm{d}r} = \pm \frac{1}{NKb} \bigg(\frac{1}{b^2}-\frac{N^2}{r^2}\bigg)^{-1/2},
\end{equation}
which leads to
\begin{eqnarray}
  \label{}
  t(r,d) & \equiv & \frac{1}{c} \int_d^r \frac{\mathrm{d}r}{NKb} \bigg(\frac{1}{b^2}-\frac{N^2}{r^2}\bigg)^{-1/2}\nonumber\\
  & = & \frac{1}{c}\sqrt{r^2-d^2}\nonumber\\
  & & +\frac{GM}{c^3}\sqrt{\frac{r-d}{r+d}} +2\frac{GM}{c^3}\ln\bigg(\frac{r+\sqrt{r^2-d^2}}{d}\bigg)\nonumber\\
  & & +\frac{\Lambda}{18c}\sqrt{r^2-d^2}(2r^2+d^2)\nonumber\\
  & & +\Lambda\frac{GM}{6c^3}\sqrt{r^2-d^2}\bigg(4r+\frac{d^2}{d+r}\bigg)\nonumber\\
  & & +\frac{13\alpha}{cd}\arccos\bigg(\frac{d}{r}\bigg).
\end{eqnarray}

In the case of superior conjunction (SC) when the receiver is on the opposite side of the Sun as seen from the emitter, by making use of conditions $r_E \gg d$ and $r_{R} \gg d$, where $r_E$ is the distance between the emitter and the Sun and $r_R$ is the distance between the reflector and the Sun, we have the time duration of light propagation as
\begin{eqnarray}
  \label{}
  \Delta t_{\mathrm{SC}} & = & 2t(r_{E},d)+2t(r_R,d)\nonumber\\
  & = & \frac{2}{c}(r_E+r_R)+4\frac{GM}{c^3}+4\frac{GM}{c^3}\ln\frac{4r_Er_R}{d^2}\nonumber\\
  & & +\frac{2\Lambda}{9c}(r_E^3+r_R^3)+\frac{26\alpha\pi}{cd}\nonumber\\
  & & +\frac{2\Lambda}{3}\frac{GM}{c^3}(2r_E^2+2r_R^2+d^2),
\end{eqnarray}
in which the last term of $\mathcal{O}(\Lambda M)$ is kept for the purpose of calculating gravitational frequency shift (see Sec. \ref{gfs}). In above one, the GR parts can return to those given by \citet{Weinberg1972Book}. The leading deviation caused by the $f(T)$ gravity is
\begin{eqnarray}
  \label{dGRtSC}
  \delta^{\mathrm{GR}}_{\Delta t_{\mathrm{SC}}} & \equiv & |\Delta t^{\mathrm{obs}}_{\mathrm{SC}} - \Delta t^{\mathrm{GR}}_{\mathrm{SC}}| \nonumber\\
  & = & +\frac{2\Lambda}{9c}(r_E^3+r_R^3) +\frac{26\alpha\pi}{cd}.
\end{eqnarray}

Similarly, at inferior conjunction (IC) when the reflector is between the emitter and the Sun and by making use of conditions $r_E \gg d$ and $r_{R} \gg d$ again, we obtain
\begin{eqnarray}
  \label{DtIC}
  \Delta t_{\mathrm{IC}} & = & 2t(r_{E},d)-2t(r_R,d)\nonumber\\
  & = & \frac{2}{c}(r_E-r_R) +4\frac{GM}{c^3}\ln\bigg(\frac{r_E}{r_R}\bigg)\nonumber\\
  & & +\frac{2\Lambda}{9c}(r_E^3-r_R^3),
\end{eqnarray}
where $d$ is canceled out due to the minus sign in the expression of $\Delta t_{\mathrm{IC}}$ and its GR parts (the first two terms) match those given by \citet{Nelson2011}. We can also have
\begin{equation}
  \label{dGRtIC}
  \delta^{\mathrm{GR}}_{\Delta t_{\mathrm{IC}}}  \equiv  |\Delta t^{\mathrm{obs}}_{\mathrm{IC}} - \Delta t^{\mathrm{GR}}_{\mathrm{IC}}| = \bigg| \frac{2\Lambda}{9c}(r_E^3-r_R^3) \bigg|.
\end{equation}
On the contrary to the light bending, the time delay experiments during IC can provide a clean constraint on $\Lambda$ regardless of $\alpha$.

\subsection{Gravitational frequency shift}

\label{gfs}

In the \textit{Cassini} experiment \citep{Bertotti2003Nature425.374}, what is measured is not the time delay but the relative change in the frequency. Around SC, a ground station transmits a radio wave signal with the frequency $\nu_0$ to the spacecraft. This signal is coherently transponded by the spacecraft and sent back to the Earth. The two-way fractional frequency fluctuation is
\begin{equation}
   \label{}
   y_{\mathrm{SC}} =  \frac{\nu(t)-\nu_0}{\nu_0} = \frac{\mathrm{d}\Delta t_{\mathrm{SC}}}{\mathrm{d}t},
\end{equation}
where the contribution due to the $f(T)$ gravity is
\begin{equation}
   \label{}
   y^{\mathrm{T}}_{\mathrm{SC}}  =  \frac{\mathrm{d}\Delta t^{\mathrm{T}}_{\mathrm{SC}}}{\mathrm{d}t} = \frac{4}{3}\Lambda \frac{GM}{c^3}d\frac{\mathrm{d}}{\mathrm{d}t}d(t) -\frac{26\alpha\pi}{cd^2}\frac{\mathrm{d}}{\mathrm{d}t}d(t),
\end{equation}
where for the \textit{Cassini} conjunction experiment $\mathrm{d}d(t)/\mathrm{d}t$ is approximately the orbital velocity of the Earth $v_{\oplus}$. The experiment starts from 12 days before the SC and ends at 12 days after it. In one day, the distance of closet approach of the signal changes by about $1.5R_{\odot}$. So the possible deviation from GR in the \textit{Cassini} experiment is
\begin{eqnarray}
  \label{dGRySC}
  \delta^{\mathrm{GR}}_{y_{\mathrm{SC}}} & \equiv & |y^{\mathrm{obs}}_{\mathrm{SC}}-y^{\mathrm{GR}}_{\mathrm{SC}}| = |y^{\mathrm{T}}_{\mathrm{SC}}(12\mathrm{d})-y^{\mathrm{T}}_{\mathrm{SC}}(0)|\nonumber\\
  & \approx & \bigg| 16 \Lambda \frac{GM_{\odot}}{c^3}R_{\odot}v_{\oplus} +\frac{8320}{729}\alpha \frac{ \pi v_{\oplus}}{cR_{\odot}^2}\bigg|,
\end{eqnarray}
which can go back to the result given by \citet{Kagramanova2006PLB634.465} when $\alpha$ is zero.

\section{Upper-bounds on $\alpha$ and $\Lambda$}

\label{bounds}

In this section we will apply above results to estimate the upper-bounds on $\alpha$ and $\Lambda$.

\subsection{Perihelion advance}

For inner planets in the Solar System, their $\delta^{\mathrm{GR}}_{\left<\dot{\omega}\right>}$ range from several tens to hundreds micro-arcseconds ($\mu$as) per century \citep{Nordtvedt2000PRD61.122001,Pitjeva2005AstL31.340,Fienga2011CMDA111.363}. Based on equation (\ref{dGRomega}) and making $\delta^{\mathrm{GR}}_{\left<\dot{\omega}\right>}$ and $a$ be expressed in the units of $10\; \mu\mathrm{as}\; \mathrm{cy}^{-1}$ and the astronomical unit (au) \footnote{We use lower-case ``au'' to represent the astronomical unit, according to International Astronomical Union 2012 Resolution B2: \url{http://www.iau.org/static/resolutions/IAU2012_English.pdf}}, we can obtain a bound as
\begin{eqnarray}
  \label{qdGRomega}
    -\frac{\delta^{\mathrm{GR}}_{\left<\dot{\omega}\right>}}{10\; \mu\mathrm{as}\; \mathrm{cy}^{-1}} & \le & + 14.7 \tilde{\Lambda}\bigg(\frac{a}{\mathrm{au}}\bigg)^{3/2}(1-e^2)^{1/2} \nonumber\\
     & & + 1.76 \times 10^3 \tilde{\alpha} \bigg(\frac{a}{\mathrm{au}}\bigg)^{-5/2} (1-e^2)  \nonumber\\
     & \le & \frac{\delta^{\mathrm{GR}}_{\left<\dot{\omega}\right>}}{10\; \mu\mathrm{as}\; \mathrm{cy}^{-1}},
\end{eqnarray}
where, to easily compare our results with those given by \citet{Iorio2012MNRAS427.1555}, based on their values, we rescale the model parameters as
\begin{equation}
  \label{}
  \tilde{\alpha}  \equiv \frac{\alpha}{10^{4}\;\mathrm{m}^2},
\end{equation}
and
\begin{equation}
  \label{}
  \tilde{\Lambda}  \equiv \frac{\Lambda}{10^{-42}\;\mathrm{m}^{-2}}.
\end{equation}
In equation (\ref{qdGRomega}), we have not taken any data yet from the planets and just do the rescaling for convenience in the calculation of estimation.

\subsection{Light bending}

In astrometric observation for gravitational light bending, the Very Long Baseline Array (VLBA) demonstrated the accuracy of measuring relative positions of radio sources can reach $\sim 10$ to 100 $\mu$as \citep{Fomalont2003ApJ598.704,Fomalont2009ApJ699.1395}, which makes us have, from equation (\ref{dGRphi}),
\begin{equation}
  \label{qdGRphi}
   8.70 \times 10^{-3} |\tilde{\alpha}| \bigg(\frac{d}{R_{\odot}}\bigg)^{-2} \le \frac{\delta^{\mathrm{GR}}_{\Delta \phi}}{10\; \mu\mathrm{as}},
\end{equation}
which is independent on $\tilde{\Lambda}$.

\subsection{Gravitational time delay}

For the gravitational time delay experiments at SC, the time delay measurements are highly dominated by the perturbations of solar corona \citep[e.g.][]{Verma2012COSPAR39.2081} and it is very difficult to separate these effects from others. Therefore, any constraints from time delay measurements should be obtained in consideration of the solar corona model for data analysis. However, the situation at IC is totally different. The biggest uncertainty does not come from the solar corona but from the positions of the receiver and the emitter. So, we only take experiments at IC into account here.

We assume the receiver is carried by a spacecraft and the emitter is on the Earth. The uncertainties of the receiver's and emitter's positions might respectively be several centimeters in the best case relative to the Earth and a few kilometers with respect to the Sun. By propagating these uncertainties in equation (\ref{DtIC}), its leading effect in light time is about $10^4$ nano-second (ns). It can impose a unique bound on $\Lambda$ as
\begin{equation}
  \label{qdGRtIC}
 \bigg| 2.48 \times 10^{-12} \tilde{\Lambda} \bigg(\frac{r_E}{\mathrm{au}}\bigg)^3 \bigg[1-\bigg(\frac{r_R}{r_E}\bigg)^3\bigg] \bigg| \le  \frac{\delta^{\mathrm{GR}}_{\Delta t_{\mathrm{IC}}}}{10^4\;\mathrm{ns}},
\end{equation}
which is independent on $\tilde{\alpha}$.

\subsection{Gravitational frequency shift}

In the \textit{Cassini} SC experiment \citep{Bertotti2003Nature425.374}, the deviation from GR is $\delta^{\mathrm{GR}}_{y_{\mathrm{SC}}} \le 10^{-14}$. This result is obtained by using two different wavelengths, which can disentangle the signals from solar corona effects. However, in most situations, it is not the case. That is the reason why the time delay experiments can not be so clearly corrected from the perturbations of solar plasma. From equation (\ref{dGRySC}), we have
\begin{equation}
  \label{qdGRySC}
   | 1.65 \times 10^{-19} \tilde{\Lambda} + 7.41 \times 10^{-3} \tilde{\alpha} | \le  \frac{\delta^{\mathrm{GR}}_{y_{\mathrm{SC}}}}{10^{-14}}.
\end{equation}

\subsection{Upper-bounds}

By solving the linear system about $\tilde{\alpha}$ and $\tilde{\Lambda}$ consisting of six inequalities generated by equation (\ref{qdGRomega}) based on supplementary advances \citep[see][Table 5]{Fienga2011CMDA111.363} in the perihelia of the planets from Mercury to Saturn with the INPOP10a ephemeris \citep{Fienga2010Journees} and four inequalities respectively given by equations (\ref{qdGRphi})-(\ref{qdGRySC}), we can estimate our upper-bounds on $\alpha$ and $\Lambda$ as
\begin{eqnarray}
  \label{}
  |\tilde{\alpha}| & \le & 1.2 \times 10^{-2},\\
  |\tilde{\Lambda}| & \le & 1.8 \times 10^{-1},
\end{eqnarray}
and, more explicitly,
\begin{eqnarray}
  \label{}
  |\alpha| & \le & 1.2 \times 10^{2}\;\mathrm{m}^2,\\
  |\Lambda| & \le & 1.8 \times 10^{-43}\;\mathrm{m}^{-2}.
\end{eqnarray}
They are consistent with those of \citet{Iorio2012MNRAS427.1555} and \citet{Kagramanova2006PLB634.465} and at least 10 times tighter than those of \citet{Iorio2012MNRAS427.1555}. They also justify the approximation that we ignore terms with the order of $\mathcal{O}(\alpha^2/r^4)$ in the spacetime given by equation (\ref{sphsym}) as \citet{Iorio2012MNRAS427.1555} did. It is worth mentioning that (i) the above upper-bounds are dominantly given by the supplementary advances \citep[see][Table 5]{Fienga2011CMDA111.363} in the perihelia of the planets. Table \ref{tab:1} shows upper-bounds given by different combinations of equations (\ref{qdGRomega}), (\ref{qdGRphi}), (\ref{qdGRtIC}) and(\ref{qdGRySC}) and clearly indicates that the supplementary advances play the most important and dominant role in our estimation; (ii) as mentioned in \citet{Fienga2010IAUS261.159}, in the construction of such supplementary advances according to observational datasets, the effects due to the Sun's quadruple mass moment $J_2^{\odot}$ are considered and isolated in the finally given results, which means they might represent possible unexplained parts of perihelion advances by GR; (iii) in our estimation, we take the allowed maximum absolute values of the supplementary advances [these values make the difference of post-fit residuals from INPOP10a below 5\% \citep{Fienga2011CMDA111.363}]; (iv) the perihelion shifts caused by the Lense-Thirring effect \citep{Lense1918PhyZ19.156} due to the Sun's angular momentum $S_{\odot}$, which is 
\begin{equation}
  \label{}
  \dot{\omega}_{\mathrm{LT}} = -\frac{6GS_{\odot}\cos i}{c^2a^3(1-e^2)^{3/2}},
\end{equation}
where $ S_{\odot} = 1.9 \times 10^{41}$ kg m${}^2$ s${}^{-1}$ \citep{Pijpers2003A&A402.683}, are included in above upper-bounds. Furthermore, all these estimations are all correlated as the estimations of the advance of perihelia obtained with INPOP10a were based on some assumptions on the Shapiro effects and a complete and rigorous test should be to implement the proposed supplementary terms in the construction of the planetary orbital solution.

\begin{table*}
 \centering
 \begin{minipage}{140mm}
  \caption{Upper-bounds obtained by different constraints.}
  \begin{tabular}{lcc}
  \hline
  Constraints  \footnote{In this table, ``PA'' denotes ``perihelion advance'' and it means the constraint given by equation (\ref{qdGRomega}) based on supplementary advances \citep[see][Table 5]{Fienga2011CMDA111.363} in the perihelia of the planets from Mercury to Saturn with the INPOP10a ephemeris \citep{Fienga2010Journees}. Similarly, ``LB'', ``TD@IC'' and ``FS@SC'' denote ``light bending'' from equation (\ref{qdGRphi}), ``time delay at IC'' from equation (\ref{qdGRtIC}) and ``frequency shift at SC'' from equation (\ref{qdGRySC}).  }      &   $|\alpha|\;(\mathrm{m}^2)$     &   $|\Lambda|\;(\mathrm{m}^{-2})$ \\
  \hline
  PA  &   $ \le 1.2 \times 10^{2} $  & $\le 1.8 \times 10^{-43}$\\
  LB & $\lesssim 3\times 10^7$ \footnote{This estimation is obtained by taking $d\sim 5R_{\odot}$ and $\delta^{\mathrm{GR}}_{\Delta\phi}\sim 10\,\mu\mathrm{as}$. } & --  \\
  TD@IC & -- & $\lesssim 6\times 10^{-30}$ \footnote{This estimation is obtained by taking $r_E\sim 1\,\mathrm{au}$, $r_R/r_E\sim0.7$ and $\delta^{\mathrm{GR}}_{\Delta t_{\mathrm{IC}}}\sim 10^4\,\mathrm{ns}$.}\\
   FS@SC \footnote{$\alpha$ and $\Lambda$ are linearly correlated in equation (\ref{qdGRySC}) and can not be separated by using equation (\ref{qdGRySC}) itself alone.} & -- & --\\
  \hline
  PA + LB &   $ \le 1.2 \times 10^{2} $  & $\le 1.8 \times 10^{-43}$\\
  PA + TD@IC &   $ \le 1.2 \times 10^{2} $  & $\le 1.8 \times 10^{-43}$\\
  PA + FS@SC  &   $ \le 1.2 \times 10^{2} $  & $\le 1.8 \times 10^{-43}$\\
  \hline
  \label{tab:1}
\end{tabular}
\end{minipage}
\end{table*}

\section{Conclusions and discussion}

\label{candd}

In this work, we investigate effects on astronomical observation and experiments conducted in the Solar System due to the $f(T)$ gravity, to extend the previous work \citep{Iorio2012MNRAS427.1555} in which perihelion advances are considered only and to find more stringent constraints on its model parameters: $\alpha$ and $\Lambda$. Up to their leading contribution, the secular evolution of planetary orbital motions, light deflection, gravitational time delay and frequency shift are calculated. 

We find qualitatively that the light deflection holds a unique bound on $\alpha$, without dependence on $\Lambda$, and the time delay experiments during inferior conjunction impose a clean constraint on $\Lambda$, regardless of $\alpha$. Based on observation and experiments, especially the supplementary advances in the perihelia of the planets from Mercury to Saturn with the INPOP10a ephemeris \citep{Fienga2011CMDA111.363}, we find the upper-bounds on those two parameters quantitatively: $|\alpha| \le  1.2 \times 10^{2}$ m${}^2$ and $ |\Lambda| \le 1.8 \times 10^{-43}$ m${}^{-2}$, at least 10 times tighter than previous results \citep{Iorio2012MNRAS427.1555}.

We also find that modern high-precision ephemerides, such as INPOP10a \citep{Fienga2010Journees}, can serve as a very sharp tool for testing fundamental theories of gravity \textbf{(see Table \ref{tab:1})}.

Several open issues remains in testing the $f(T)$ gravity. One of them is about its effects on small scales, for example its possible deviations from GR in the vicinity of the Earth. Since terms of $\mathcal{O}(\alpha^2/r^4)$ are neglected in both the previous work \citep{Iorio2012MNRAS427.1555} and this one, the spacetime (\ref{sphsym}) is not suitable for doing this and a more general solution of spacetime in the framework of the $f(T)$ gravity is needed. After that, then it will be possible to test the $f(T)$ gravity by tracking a drag-free satellite with laser ranging or the global positioning system \citep{Damour1994PRD50.2381} and by a very precise clock onboard a drag-free satellite with laser ranging and time transfer link \citep{Deng2013MNAS431.3236}.

\section*{Acknowledgments}

We acknowledge very useful and helpful comments and suggestions from our anonymous referee. The work of YX is supported by the National Natural Science Foundation of China Grant No. 11103010, the Fundamental Research Program of Jiangsu Province of China Grant No. BK2011553, the Research Fund for the Doctoral Program of Higher Education of China Grant No. 20110091120003 and the Fundamental Research Funds for the Central Universities No. 1107020116. The work of XMD is funded by the Natural Science Foundation of China under Grant No. 11103085. This project/publication was made possible through the support of a grant from the John Templeton Foundation. The opinions expressed in this publication are those of the authors and do not necessarily reflect the views of the John Templeton Foundation. The funds from the John Templeton Foundation were provided by a grant to The University of Chicago which also managed the program in conjunction with National Astronomical Observatories, Chinese Academy of Sciences. XMD appreciates the support from the group of Almanac and Astronomical Reference Systems in the Purple Mountain Observatory of China.

\bibliographystyle{mn2e.bst}
\bibliography{Gravity20131203.bib}

\begin{thebibliography}{}

\bibitem[\protect\citeauthoryear{{Amendola} \& {Tsujikawa}}{{Amendola} \&
  {Tsujikawa}}{2010}]{Amendola2010}
{Amendola} L.,  {Tsujikawa} S.,  2010, {Dark Energy: Theory and Observations}.
{Cambridge University Press}, {Cambridge}

\bibitem[\protect\citeauthoryear{{Bengochea}}{{Bengochea}}{2011}]{Bengochea2011PLB695.405}
{Bengochea} G.~R.,  2011, Physics Letters B, 695, 405

\bibitem[\protect\citeauthoryear{{Bengochea} \& {Ferraro}}{{Bengochea} \&
  {Ferraro}}{2009}]{Bengochea2009PRD79.124019}
{Bengochea} G.~R.,  {Ferraro} R.,  2009, \prd, 79, 124019

\bibitem[\protect\citeauthoryear{{Bertotti}, {Iess} \& {Tortora}}{{Bertotti}
  et~al.}{2003}]{Bertotti2003Nature425.374}
{Bertotti} B.,  {Iess} L.,    {Tortora} P.,  2003, \nat, 425, 374

\bibitem[\protect\citeauthoryear{{Clifton}, {Ferreira}, {Padilla} \&
  {Skordis}}{{Clifton} et~al.}{2012}]{Clifton2012PhR513.1}
{Clifton} T.,  {Ferreira} P.~G.,  {Padilla} A.,    {Skordis} C.,  2012,
  \physrep, 513, 1

\bibitem[\protect\citeauthoryear{{Damour} \& {Esposito-Far{\`e}se}}{{Damour} \&
  {Esposito-Far{\`e}se}}{1994}]{Damour1994PRD50.2381}
{Damour} T.,  {Esposito-Far{\`e}se} G.,  1994, \prd, 50, 2381

\bibitem[\protect\citeauthoryear{{Danby}}{{Danby}}{1962}]{Danby1962}
{Danby} J.,  1962, {Fundamentals of Celestial Mechanics}.
Macmillan, New York

\bibitem[\protect\citeauthoryear{{de Felice} \& {Tsujikawa}}{{de Felice} \&
  {Tsujikawa}}{2010}]{Felice2010LRR13.3}
{de Felice} A.,  {Tsujikawa} S.,  2010, Living Reviews in Relativity, 13, 3

\bibitem[\protect\citeauthoryear{Deng \& Xie}{Deng \&
  Xie}{2013}]{Deng2013MNAS431.3236}
Deng X.-M.,  Xie Y.,  2013, \mnras, 431, 3236

\bibitem[\protect\citeauthoryear{{Ferraro} \& {Fiorini}}{{Ferraro} \&
  {Fiorini}}{2007}]{Ferraro2007PRD75.084031}
{Ferraro} R.,  {Fiorini} F.,  2007, \prd, 75, 084031

\bibitem[\protect\citeauthoryear{{Ferraro} \& {Fiorini}}{{Ferraro} \&
  {Fiorini}}{2008}]{Ferraro2008PRD78.124019}
{Ferraro} R.,  {Fiorini} F.,  2008, \prd, 78, 124019

\bibitem[\protect\citeauthoryear{{Fienga}, {Laskar}, {Kuchynka}, {Le
  Poncin-Lafitte}, {Manche} \& {Gastineau}}{{Fienga}
  et~al.}{2010}]{Fienga2010IAUS261.159}
{Fienga} A.,  {Laskar} J.,  {Kuchynka} P.,  {Le Poncin-Lafitte} C.,  {Manche}
  H.,    {Gastineau} M.,  2010, in {Klioner} S.~A.,  {Seidelmann} P.~K.,
  {Soffel} M.~H.,  eds, IAU Symposium Vol.~261 of IAU Symposium, {Gravity tests
  with INPOP planetary ephemerides}.
pp 159--169

\bibitem[\protect\citeauthoryear{{Fienga}, {Laskar}, {Kuchynka}, {Manche},
  {Desvignes}, {Gastineau}, {Cognard} \& {Theureau}}{{Fienga}
  et~al.}{2011}]{Fienga2011CMDA111.363}
{Fienga} A.,  {Laskar} J.,  {Kuchynka} P.,  {Manche} H.,  {Desvignes} G.,
  {Gastineau} M.,  {Cognard} I.,    {Theureau} G.,  2011, Celestial Mechanics
  and Dynamical Astronomy, 111, 363

\bibitem[\protect\citeauthoryear{{Fienga}, {Manche}, {Kuchynka}, {Laskar} \&
  {Gastineau}}{{Fienga} et~al.}{2010}]{Fienga2010Journees}
{Fienga} A.,  {Manche} H.,  {Kuchynka} P.,  {Laskar} J.,    {Gastineau} M.,
  2010, in {Capitaine} N.,  ed., Proceedings of the Journ\'ees 2010
  ``Syst\`emes de R\'ef\'erence Spatio-Temporels" \mbox{}, {Planetary and Lunar
  ephemerides, INPOP10a}.
pp 37--42

\bibitem[\protect\citeauthoryear{{Fomalont}, {Kopeikin}, {Lanyi} \&
  {Benson}}{{Fomalont} et~al.}{2009}]{Fomalont2009ApJ699.1395}
{Fomalont} E.,  {Kopeikin} S.,  {Lanyi} G.,    {Benson} J.,  2009, \apj, 699,
  1395

\bibitem[\protect\citeauthoryear{{Fomalont} \& {Kopeikin}}{{Fomalont} \&
  {Kopeikin}}{2003}]{Fomalont2003ApJ598.704}
{Fomalont} E.~B.,  {Kopeikin} S.~M.,  2003, \apj, 598, 704

\bibitem[\protect\citeauthoryear{{Hayashi} \& {Shirafuji}}{{Hayashi} \&
  {Shirafuji}}{1979}]{Hayashi1979PRD19.3524}
{Hayashi} K.,  {Shirafuji} T.,  1979, \prd, 19, 3524

\bibitem[\protect\citeauthoryear{{Iorio} \& {Saridakis}}{{Iorio} \&
  {Saridakis}}{2012}]{Iorio2012MNRAS427.1555}
{Iorio} L.,  {Saridakis} E.~N.,  2012, \mnras, 427, 1555

\bibitem[\protect\citeauthoryear{{Kagramanova}, {Kunz} \&
  {L{\"a}mmerzahl}}{{Kagramanova} et~al.}{2006}]{Kagramanova2006PLB634.465}
{Kagramanova} V.,  {Kunz} J.,    {L{\"a}mmerzahl} C.,  2006, Physics Letters B,
  634, 465

\bibitem[\protect\citeauthoryear{Lake}{Lake}{2002}]{Lake2002PRD65.087301}
Lake K.,  2002, Phys. Rev. D, 65, 087301

\bibitem[\protect\citeauthoryear{{Lense} \& {Thirring}}{{Lense} \&
  {Thirring}}{1918}]{Lense1918PhyZ19.156}
{Lense} J.,  {Thirring} H.,  1918, Physikalische Zeitschrift, 19, 156

\bibitem[\protect\citeauthoryear{{Linder}}{{Linder}}{2010}]{Linder2010PRD81.127301}
{Linder} E.~V.,  2010, \prd, 81, 127301

\bibitem[\protect\citeauthoryear{{Nelson}}{{Nelson}}{2011}]{Nelson2011}
{Nelson} R.~A.,  2011, Metrologia, 48, 171

\bibitem[\protect\citeauthoryear{Nordtvedt}{Nordtvedt}{2000}]{Nordtvedt2000PRD61.122001}
Nordtvedt K.,  2000, Phys. Rev. D, 61, 122001

\bibitem[\protect\citeauthoryear{{Perlmutter}, {Aldering}, {Goldhaber} \& {et
  al.}}{{Perlmutter} et~al.}{1999}]{Perlmutter1999ApJ517.565}
{Perlmutter} S.,  {Aldering} G.,  {Goldhaber} G.,    {et al.} 1999, \apj, 517,
  565

\bibitem[\protect\citeauthoryear{{Pijpers}}{{Pijpers}}{2003}]{Pijpers2003A&A402.683}
{Pijpers} F.~P.,  2003, \aap, 402, 683

\bibitem[\protect\citeauthoryear{{Pitjeva}}{{Pitjeva}}{2005}]{Pitjeva2005AstL31.340}
{Pitjeva} E.~V.,  2005, Astronomy Letters, 31, 340

\bibitem[\protect\citeauthoryear{{Riess}, {Filippenko}, {Challis} \& {et
  al.}}{{Riess} et~al.}{1998}]{Riess1998AJ116.1009}
{Riess} A.~G.,  {Filippenko} A.~V.,  {Challis} P.,    {et al.} 1998, \aj, 116,
  1009

\bibitem[\protect\citeauthoryear{{Unzicker} \& {Case}}{{Unzicker} \&
  {Case}}{2005}]{Unzicker2005}
{Unzicker} A.,  {Case} T.,  2005, arXiv:physics/0503046

\bibitem[\protect\citeauthoryear{{Verma} \& {Fienga}}{{Verma} \&
  {Fienga}}{2012}]{Verma2012COSPAR39.2081}
{Verma} A.,  {Fienga} A.,  2012, in 39th COSPAR Scientific Assembly Vol.~39 of
  COSPAR Meeting, {Electron density distribution and solar plasma correction of
  radio signals using MGS, MEX and VEX spacecraft navigation data}.
p.~2081

\bibitem[\protect\citeauthoryear{{Weinberg}}{{Weinberg}}{1972}]{Weinberg1972Book}
{Weinberg} S.,  1972, {Gravitation and Cosmology: Principles and Applications
  of the General Theory of Relativity}.
{John Wiley \& Sons, Inc.}, {New York London Sydney Toronto}

\bibitem[\protect\citeauthoryear{{Wu} \& {Yu}}{{Wu} \&
  {Yu}}{2010}]{Wu2010PLB693.415}
{Wu} P.,  {Yu} H.,  2010, Physics Letters B, 693, 415

\end{thebibliography}

\end{document}